\documentclass[12pt]{article}

\usepackage{amsmath,amssymb,pifont}
\usepackage{graphicx,color,subfigure,bm,rotating}
\usepackage{booktabs}

\newenvironment{sciabstract}{%
\begin{quote} \bf}
{\end{quote}}

\newcounter{lastnote}

\title{Bubbles determine the amount of alcohol in Mezcal}

\author{G. Rage,$^{1}$ O. Atasi,$^{2,3}$ M. M. Wilhelmus ,$^{4}$ J. F. Hern\'andez-S\'anchez,$^{5}$  \\ B. Haut,$^2$ B. Scheid,$^2$ D. Legendre,$^3$ and R. Zenit$^{1,\ast}$
\\
\normalsize{$^{1}$Instituto de Investigaciones en Materiales, Universidad Nacional Aut\'onoma de M\'exico}\\
\normalsize{Ciudad de M\'exico, 04510, Mexico}\\
\normalsize{$^{2}$TIPs (Transfers, Interfaces and Processes), Universit\'e libre de Bruxelles}\\
\normalsize{Avenue F.D. Roosevelt, 50, CP 165/67, 1050 Brussels, Belgium}\\
\normalsize{$^{3}$Institut de M\'ecanique des Fluides de Toulouse, Universit\'e de Toulouse}\\
\normalsize{CNRS, Toulouse, France}\\
\normalsize{$^{4}$Department of Mechanical Engineering, University of California, Riverside} \\
\normalsize{3401 Watkins Drive, Bourns Hall A313, Riverside, CA 92521, USA}\\
\normalsize{$^{5}$Division of Physical Sciences and Engineering}\\
\normalsize{King Abdullah University of Science and Technology}\\
\normalsize{Thuwal, 23955-6900, Saudi Arabia}\\
\normalsize{$^\ast$To whom correspondence should be addressed; e-mail:  zenit@unam.mx}
}

\date{}

\begin{document}


\baselineskip24pt

\maketitle

\begin{sciabstract}
Mezcal is a traditional alcoholic Mexican spirit distilled from fermented agave juices that has been produced for centuries. Its preparation and testing involves an artisanal method to determine the alcohol content based on pouring a stream of the liquid into a small vessel: if the alcohol content is correct, stable bubbles, known as pearls, form at the surface and remain floating for some time. It has been hypothesized that an increase in bubble lifetime results from a decrease in surface tension due to added surfactants. However, the precise mechanism for extended lifetime remains unexplained. By conducting experiments and numerical simulations, we studied the extended lifetime of pearls. It was found that both changes in fluid properties (resulting from mixing ethanol and water) and the presence of surfactants are needed to observe pearls with a long lifetime. Moreover, we found that the dimensionless lifetime of a bubble first increases with the Bond number, until reaching a maximum at $Bo\approx 1$, and then continuously decreases. Our findings on bubble stability in Mezcal not only explain the effectiveness of the artisanal method, but it also provides insight to other fields where floating bubbles are relevant such as in oceanic foam, bio-foams, froth flotation and magma flows.
\end{sciabstract}

Drinking is an essential human activity. The consumption of alcoholic drinks became a necessity for ancient humans, since it represented a reliable source of potable liquids\cite{Patrick1970}. As such, these drinks became part of the cultural identity \cite{Gately2008}: most cultures around the world have a `local' drink. In the case of Mexico, the most popular distilled spirit is Tequila which, in fact, belongs to a wider class of distilled agave-based products that are similar in production  \cite{Cedeno:1995} but vary regionally. It was believed that the production of agave-distilled spirits began with the arrival of Europeans by the end of the 16th century\cite{Colunga:2007}; however, recent findings indicate that alcohol distillation was known in Mesoamerica long before, for at least 25 centuries\cite{Goguitchaichvili2018}.
The present study is focused on Mezcal, which has progressively gained world-wide recognition. Although its production and denomination are normed \cite{NOM070}, it is mostly prepared in an artisanal manner\cite{Lopez:2010}.

\begin{figure}
\centering
\subfigure[]{\includegraphics[height=0.4\textwidth]{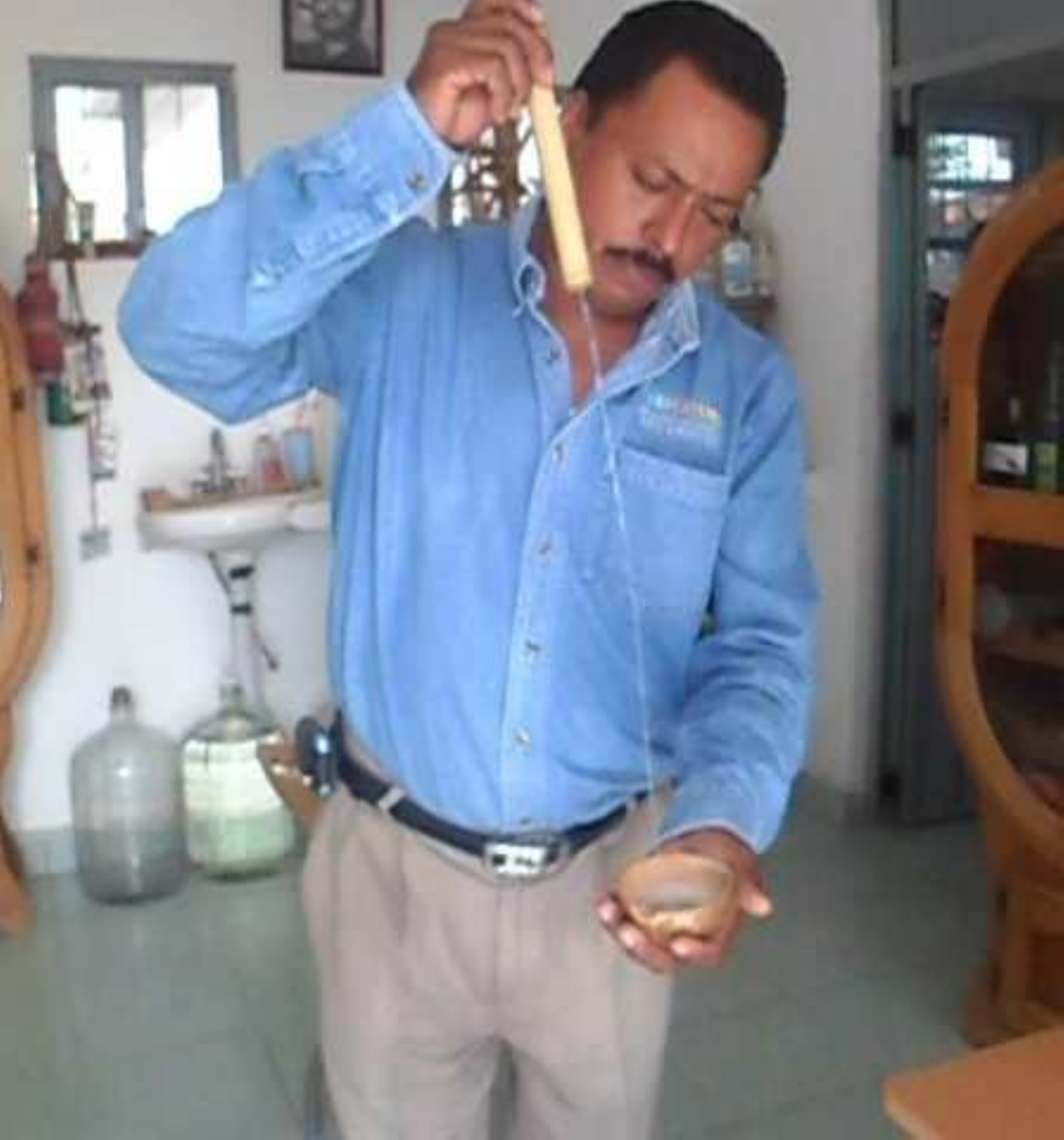}}
\subfigure[]{\includegraphics[height=0.4\textwidth]{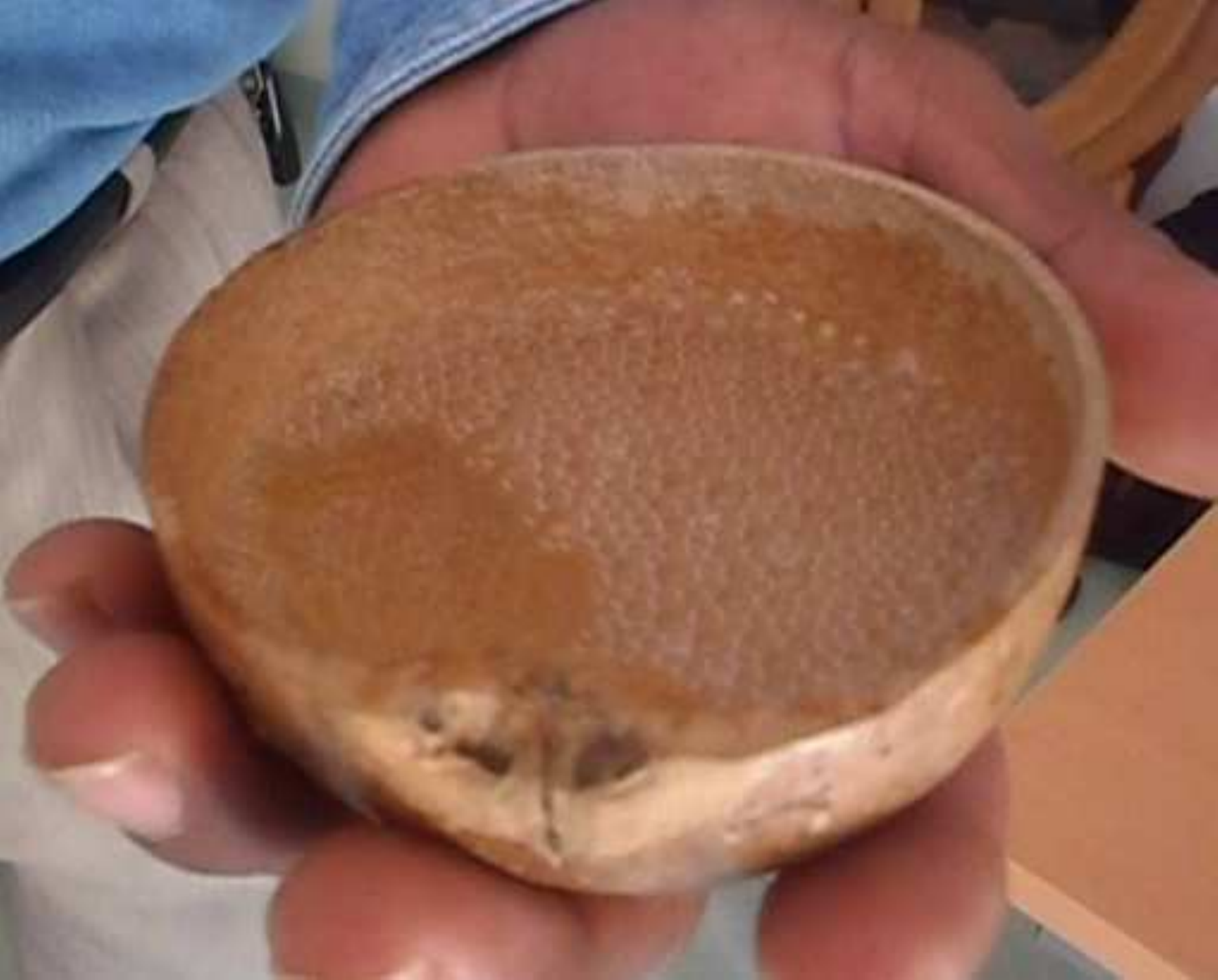}}
\subfigure[]{\includegraphics[width=0.95\textwidth]{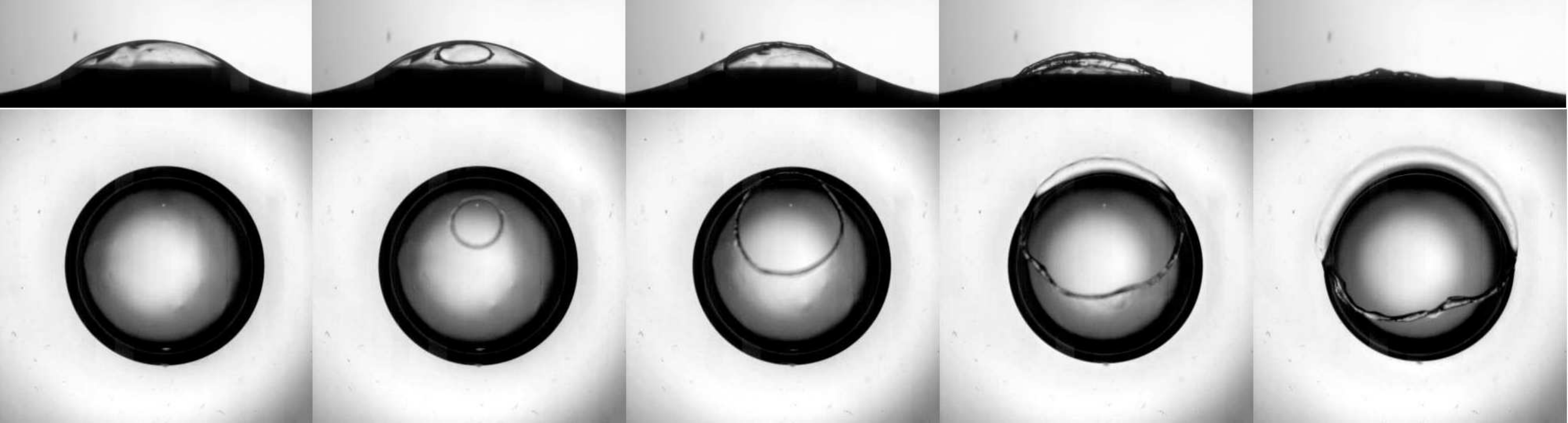}}
\caption{(a) Traditional technique to form superficial bubbles (pearls). (b) Zoom of the pearls of Mezcal in a gourd cup (\emph{j\'icara}). (c) The bursting of a surface bubble. Images taken from the side and top, simultaneously. The time between frames is 0.4 ms. The diameter of the bubble is 1.9 mm, the liquid is Mezcal M1, from Table 1.SM , corresponding to $Bo=1.06$.}
\label{fig:formation}
\end{figure}

Our interest arises from the traditional method employed to determine alcohol content in Mezcal.  According to popular accounts (informally documented by interviews with artisans) and a few formal reports\cite{Bowen2015}, the method consists of observing the lifetime of so-called pearls, bubbles that are formed by splashing a jet of fluid into a small container, see Fig. \ref{fig:formation}(a). As a result of the continuous splashing, the liquid surface breaks and bubbles form. If the amount of alcohol in the liquid is correct (about 55\% volume fraction of ethanol) pearls persist for up to tens of seconds, see Fig.\ref{fig:formation}(b). The method, described in the Methods Section,  is surprisingly accurate. Interestingly, a similar technique has also been used to determine the alcoholic content in other spirits. Davidson \cite{Davidson:1981}, for example, conducted experiments on the foam stability of Bourbon diluted with different amounts of water. The technique is essentially the same as that presented here and shows the same phenomena: for volumetric contents of alcohol of about 50\%, the superficial bubbles are notably more durable than in other mixtures. Ahmed and Dickinson\cite{Ahmed:1990} conducted experiments for whiskey and found similar results. They argue that the changes in lifetime are related to the changes in solubility of surfactant molecules present in this type of beverages. In contrast, Tuinier et al. \cite{Tuinier1996} found an extended foam-life for ethanol volume fractions close to 10\%. In none of the previous studies the precise mechanism responsible for the extended bubble life duration has been explained.

Due to its relevance to many processes, e.g. bubbles on the surface of the ocean\cite{Blanchard:1957,Garrett:1967}, fish nests \cite{Jaro2001}, champagne\cite{Liger2009}, froth floatation\cite{Shean2011} and vulcanology\cite{Manga1994}, the time of residence of a superficial bubble on a free surface has been extensively studied. How long a bubble remains floating on a surface depends on the drainage of its film, which results from the balance between two competing effects (gravitational and capillary induced drainage) and viscous forces. When the film is sufficiently thin, it spontaneously pierces and breaks. Therefore, we can define a dimensionless bubble lifetime, $T^{**}_{life}$ as (see details in the Appendices)
\begin{equation}\label{eqn:Tlife}
  T^{**}_{life}=\frac{T_{life}\sigma h_{rup}}{\mu D^3}
\end{equation}
where $\mu$ is the fluid viscosity, $\sigma$ is the surface tension, $h_{rup}$ is the rupture film thickness and $D$ is the bubble diameter. The ratio of the two competing draining effects is the Bond number, defined as:
\begin{equation}\label{eqn:Bond}
  Bo=\frac{\rho g D^2}{\sigma}
\end{equation}
where $\rho$ is the liquid density and $g$ the gravitational acceleration. In the Appendices we provide a physical interpretation of the Bond number.

In addition to the physical variables discussed above, it has been long recognized that the presence of surfactants significantly alters the draining process of the bubble film, thereby affecting its lifetime. Of particular interest to the present study is the effect of alcohol, which is not fully understood yet as the effect of evaporation can shorten the lifetime of films \cite{Hosoi:2001}, or extend it \cite{Gordeeva:2013}. Indeed,  some studies suggest that alcohol can increase bubble stability\cite{Ahmed:1990}, others have shown it to have a destabilizing effect\cite{Edmonstone:2004}.

In Mezcal, water and ethanol are the main components. The ethanol volume fraction ranges from 36 to 55 \%, according to the norm\cite{NOM070}. However, and perhaps more importantly, many other components are contained in small fractions\cite{Leon:2006}: methanol, acetic acid, ethyl acetate, higher alcohols, esters, ketones, furanes, acetals, aldehydes, phenols and terpenes, among others. Some of these may act as surfactants.
Additionally, proteins, possibly present in Mezcal, can have a significant effect in delaying the draining of bubble films, as in the case of beer\cite{Blasco:2011}.

It is beyond the scope of the present study to explain how the traditional technique to assess the ethanol content originated. What we aim to explain in this investigation is the relation between the lifetime of pearls and alcohol content. Although modern methods of determining alcohol content are accessible and reliable, even for small remote artisanal communities, the fundamental understanding of the physical mechanisms of bubble and foam stability are of importance to many other fields.

Given that the bubble formation process is relatively well understood (see Appendices), we focus on the time that a bubble could remain floating on the liquid surface before bursting. Pearl lifetime was measured experimentally by placing a single bubble at the liquid surface of a small container and capturing its life span via video recordings (Methods Section). Figure \ref{fig:formation} (c) shows a typical bubble at the moment of bursting. It was observed that bursting initiates with a small hole rupturing the film, which opens quickly until it retracts completely.
The puncture does not typically appear at the apex of the bubble; the thinning of the film is not uniform due to the partial regeneration mechanism\cite{Lhuissier2012}.

\begin{figure}
\centering
\vspace{0.5cm}
\subfigure[]{\includegraphics[width=0.60\textwidth]{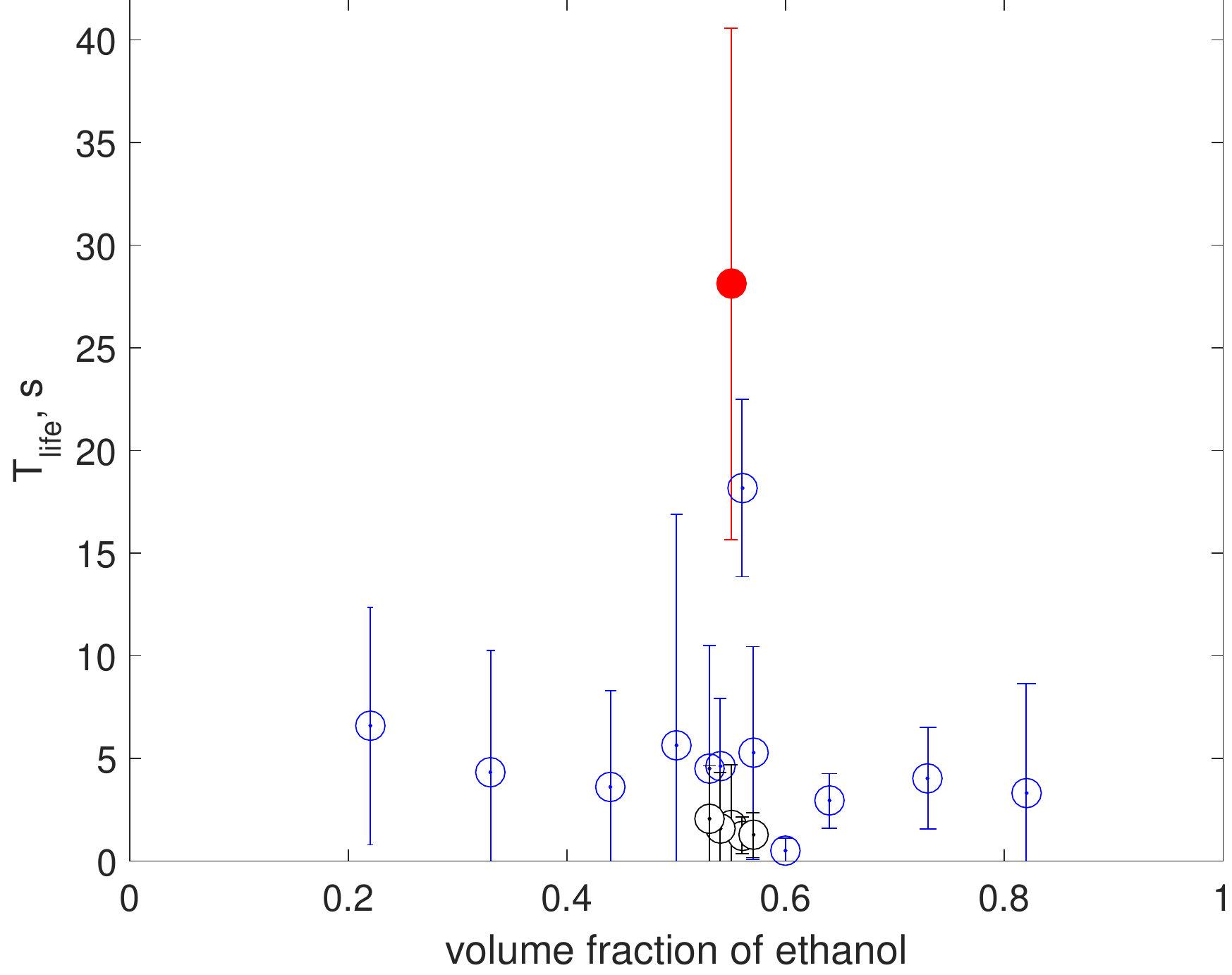}} \vspace{0.5cm}
\subfigure[]{\includegraphics[width=0.60\textwidth]{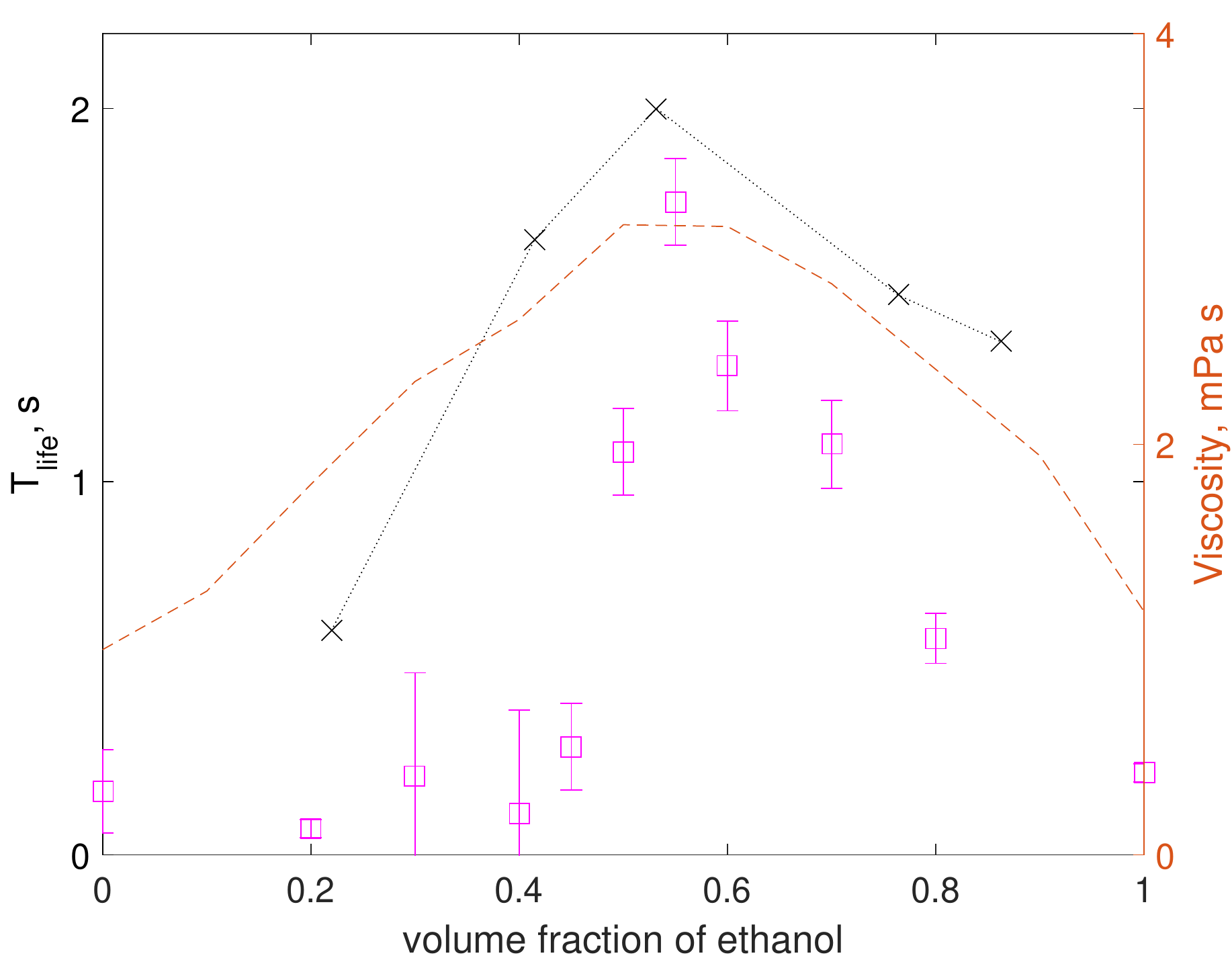}}
\caption{Bubble lifetime as a function of ethanol concentration. (a) Mezcal (M1,{\color{red}$\bullet$}), Mezcal with water and/or ethanol (MA1,{\color{blue}$\circ$}, and MA2,{\color{black}$\circ$}); symbols as in Table 1 of the Appendices. (b) Water/ethanol mixtures (WE,{\color{magenta}$\square$}, left axis); numerical results ({\color{black}$\times$}) symbols and dotted line show the trend obtained from numerical simulations (in arbitrary units), for a surfactant concentration of 0.0125 mol/l; the dashed line shows the trend of the change of viscosity of water-ethanol mixtures, according to \cite{Khattab:2012} (right axis). Error bars denote the standard deviation of the measurements.}
\label{fig:test}
\end{figure}

Experiments were initially conducted using Mezcal with the `correct' ethanol volume fraction, identified as M1 (see Table 1, Appendices). The mean lifetime in this case was 28.1 $\pm 12.5$ s. To vary the amount of ethanol in this liquid, either pure water or ethanol was added. A sharp maximum was observed for the Mezcal sample with 55\% ethanol (filled red marker in Fig. \ref{fig:test}.a). Further reduction in bubble lifetime resulted from either an increase or decrease of the volumetric fraction of alcohol in the solution.  It should be noted that even though the amount of other components was diluted (by adding water or ethanol), increasing or decreasing the alcohol content led to a measurable variation of the lifetime. Therefore, based on these observations, we conclude that the traditional technique does work to detect the correct amount of ethanol in Mezcal. We attribute the large error bars in our measurements to the use of unfiltered artisanal Mezcal. According to the norm \cite{NOM070}, particulate matter is expected to be present, which could explain the relatively large bursting thickness of the film.

To begin elucidating the underlying physical mechanism of the process we performed an experiment considering mixtures of water and ethanol, varying the volume fraction of ethanol ranging from pure water to pure ethanol, see Fig. \ref{fig:test}(b). It was observed that even when no significant amounts of other substances are present, the maximum pearl lifetime occurs at approximately 55\% of ethanol fraction. Note, however, that the lifetimes for these mixtures are one order of magnitude smaller than those observed for Mezcal. Assuming that the dimensionless bubble lifetime (Eqn. \ref{eqn:Tlife}) is only a function of the Bond number (Eqn. \ref{eqn:Bond}), we can argue that the lifetime of pearls is directly proportional to the fluid viscosity. Indeed, while density and surface tension of water-ethanol mixtures have a monotonic behavior with ethanol fraction (from the pure water to the pure ethanol values), viscosity does not: its value increases from that of water, reaching a maximum at around 55\% of ethanol, to then decrease to reach the value of pure ethanol \cite{Khattab:2012}. The dotted line in Fig. \ref{fig:test}(b) shows how viscosity varies (different scale) as the volume of ethanol increases.

To evaluate the effect of surfactants and other components, another set of experiments was conducted. In these, different amounts of a  55\% water-ethanol mixture were added to the sample M1, also shown in Fig. \ref{fig:test}(a). In this manner, the amount of ethanol remains approximately constant but the quantity of surfactants is diluted with respect to the original sample. Given that pearl lifetime was significantly reduced compared to the original sample, we can argue that both the physical properties of the liquid and the amount of surfactants are important factors to determine the lifetime of pearls.

\begin{figure}
\centering
{\includegraphics[width=0.8\textwidth]{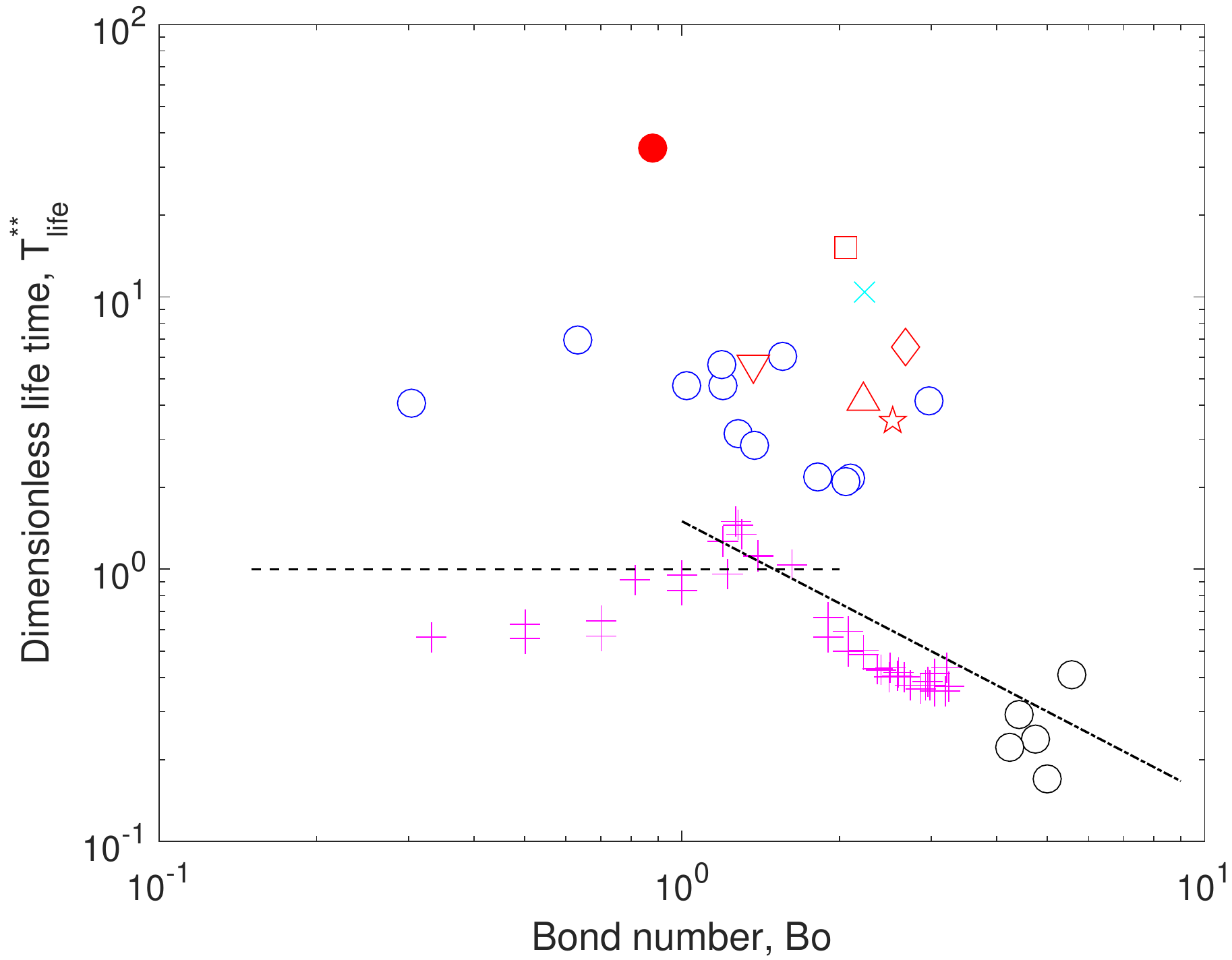}}
\caption{Dimensionless lifetime as a function of Bond number considering the viscous/capillary time scaling (Eqn.1). The ({\color{magenta}$+$}) symbols are the results of numerical simulations, for a surfactant concentration of 0.025 mol/l. Symbols as in Table 1 of the Appendices. For clarity, error bars are omitted.  The dashed line show expected trend for $Bo \ll 1$; dashed-dotted line shows the prediction from \cite{Scheid2012}, valid for $Bo \gg 1$ (see Appendices).}
\label{fig:dimensionlesslife}
\end{figure}

To understand the process in a more general manner, we present the results in dimensionless terms, considering Eqns. \ref{eqn:Tlife} and \ref{eqn:Bond}. The dimensionless bubble lifetime as a function of the Bond number is shown in Fig. \ref{fig:dimensionlesslife}, for all the samples. It is observed that the bubble lifetime in unaltered Mezcal is higher than in those samples in which either water, ethanol or water-ethanol mixture was added. More importantly, the value of $Bo$ for which the maximum time is observed is near unity. Considering the altered Mezcal cases (fluids MA1 and MA2,  Table 1.SM), the lifetime increases monotonically when $Bo<1$, while at around $Bo\approx 1$ a slight change of trend can be observed. For $Bo>1$ the lifetime  remains relatively constant. Note that the measurements conducted for other unaltered Mezcal types were also observed to follow this behavior. The Bond number allows us to evaluate the driving force for drainage, namely the capillary force for $Bo \ll 1$ and the gravity force for $Bo \gg 1$. The striking feature of the results in Fig.\ref{fig:dimensionlesslife} is that $Bo\sim 1$ indicates a transition, which coincides with the maximum bubble lifetime.

Numerical simulations of a bubble resting on a free surface with surfactants were conducted to further investigate the break up process in detail (see details in the Appendices).
Figure \ref{fig:numerical} shows the fluid velocity (tangent to the interface) as a function of position within the film, at a certain angle from the apex ($\theta=0.35$ radians) at different times,  with and without surfactants. An important difference between the two cases is that speed is significantly larger and uniform (across the thickness) for the case of a bubble without surfactants, in comparison to the case with surfactants. The reason for this reduction stems, first, from the immobilization of the surfaces due to surfactants (more prominently at $s=0$).
Moreover, when surfactants are present, for the angle shown, the fluid at the inner edge of the film ($s=0$) actually moves in the opposite direction to that of the gravitational draining, even for early times. For later times, the fluid near the outer edge ($s=1$) also moves upwards. As a result, the draining time is much longer. Note that the simulations consider the effect of surfactant transport but do not contemplate other effects (see Appendices). While we can argue that surfactant transport dominates,  the lifetimes obtained numerically are smaller than those found experimentally.  Figure \ref{fig:test}(b) shows the numerically obtained bubble lifetime, in arbitrary units, as a function of alcohol concentration. The trend is close to that obtained experimentally, for the case of water-alcohol: a clear maximum is observed at a concentration of about 55\% of alcohol. Based on these results, the maximum lifetime is clearly a result of both the physical properties of the liquid and the surfactant concentration. Figure \ref{fig:dimensionlesslife} also shows the numerical results but in dimensionless form, along with the experiments. Clearly, the numerical results show an increasing trend for small $Bo$, but at around $Bo\approx1$ a clear change of behavior is observed. The lifetime of the numerical results is shorter than that in the experiments; a detailed discussion is included in the Appendices.
\begin{figure}
\centering
\includegraphics[width=0.8\textwidth]{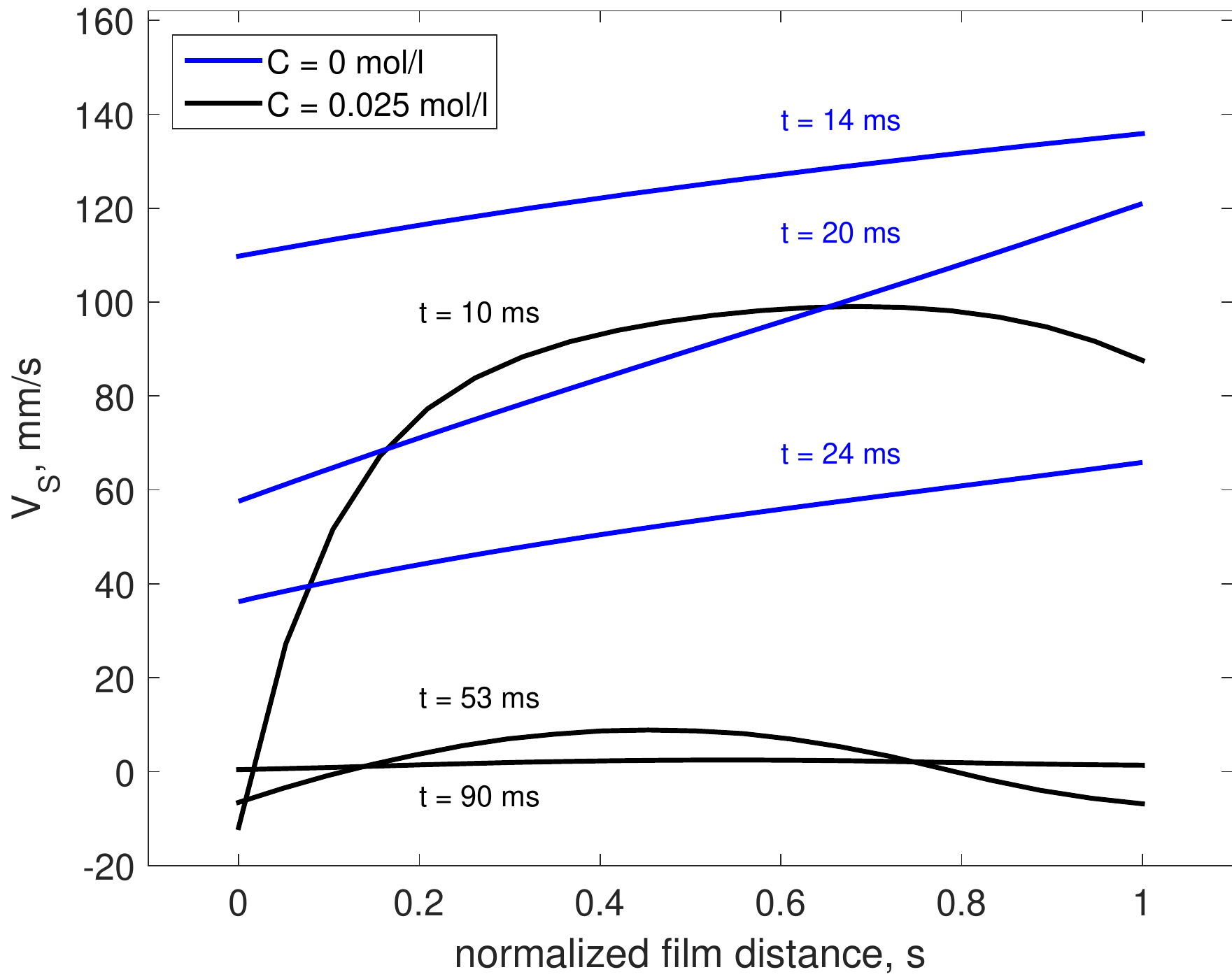}
\caption{Velocity profile within the film. Tangential fluid velocity, $V_s$, as a function of the normalized distance across the film, $s$, at 0.35 rad from the bubble apex, for different times and surfactant concentrations. $s$ = 0 corresponds to the bubble surface and $s$ = 1 refers to the external free surface. The black and blue lines correspond to the cases with and without surfactants, respectively.}
\label{fig:numerical}
\end{figure}

As discussed in the Appendices, the Bond number compares the two main driving forces for drainage: gravitational drainage for large $Bo$ and capillary-induced drainage for small $Bo$. The trends in both limits  resulting from scaling arguments are plotted in Fig. ~\ref{fig:dimensionlesslife}. The cross-over for which gravity and capillary effects are of the same order of magnitude shows a non-monotonous behaviour for which a maximum lifetime has been observed, both numerically and experimentally. Strikingly, $Bo\approx 1$ corresponds to bubble size of about 2mm, i.e. the size for which the lifetime is the longest.

Finally, to further support the existence of a critical $Bo$, we compared our results with predictions from the literature (discussed in detail in the Appendices).  These prediction also corroborate the existence of a value of $Bo$ at which $T^{*}_{life}$ has a maximum value.

We have found that the dimensionless lifetime of a surface bubble increases with Bond number, until reaching a maximum value at $Bo\approx1$, to then decreases for large values of $Bo$. It was found that both an increase of the liquid viscosity and the presence of surfactants are need to observe a long the lifetime of bubbles. These factors explain why pearls in Mezcal have a particularly long life at a certain concentration of alcohol and a certain size. Clearly, the artisanal technique is the result of observation, empiricism and tradition. The explanation of why it works was obtained by conducting controlled experiments, numerical simulations and modeling.


\section*{Acknowledgements}
R. Z. acknowledges the support of the Fulbright-Garcia Robles foundation during his sabbatical year at Caltech, where this investigation started.

\clearpage


\appendix

\section{Materials and Experimental Methods}

\subsection{Pearl formation}
The traditional technique to evaluate the alcohol content was replicated in a controlled manner. As shown in Movie \#1 (online submission) and in Fig. \ref{fig:formation}(a) , the \emph{Maestro Mezcalero} issues a stream of fluid from a long reed, with a round opening of about 2 mm in diameter. The jet impinges onto a small gourd cup, of about 10 cm in diameter. As the cup fills up, a pool of fluid is formed of several centimeters in depth. At this point, the continuous splashing of the jet breaks the surface of the liquid to form bubbles that rise to the surface, shown in Fig. 1(b) of the main document.

\begin{figure}
\centering
{\includegraphics[width=0.7\textwidth]{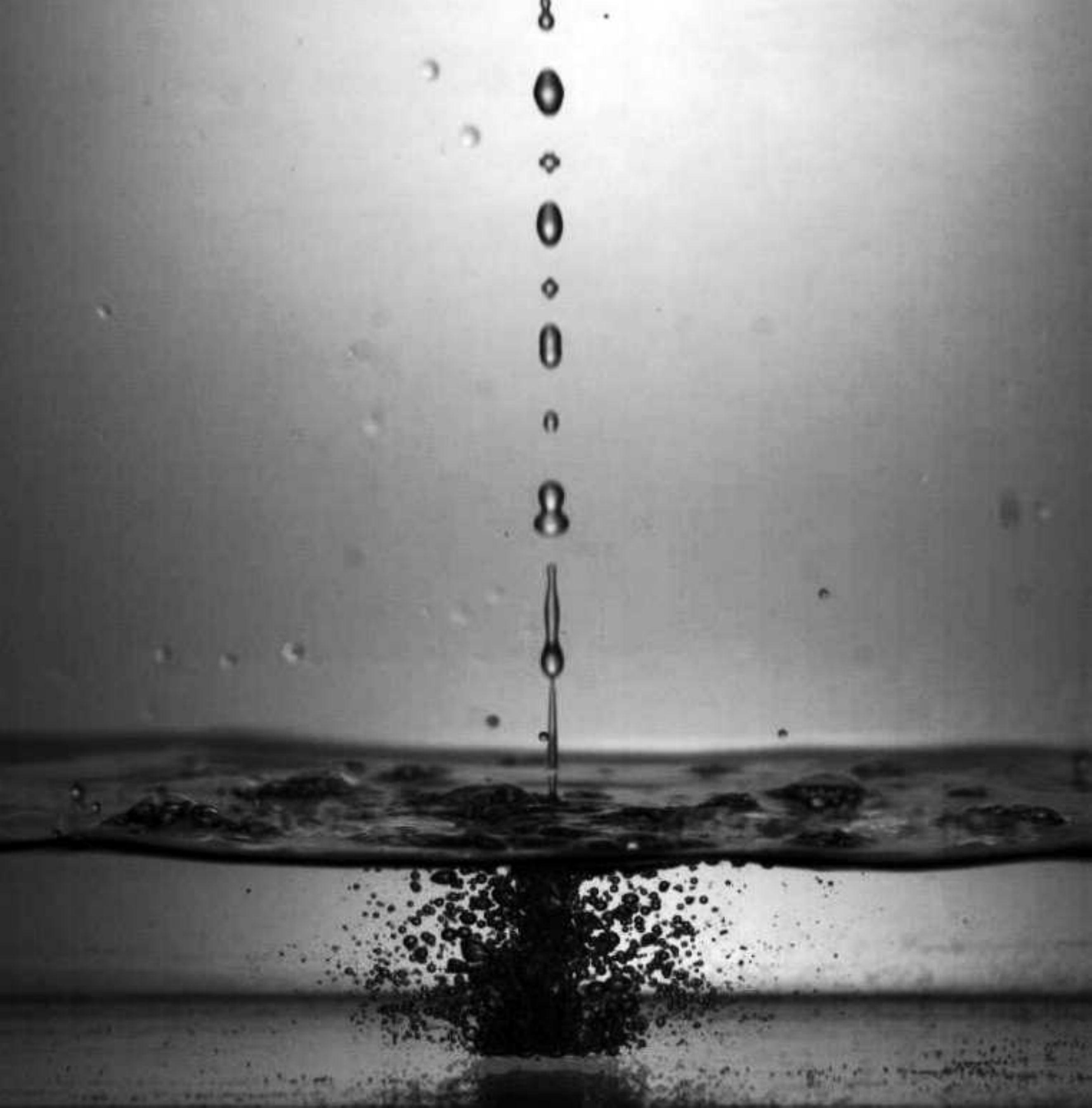}}
\caption{Reproduction of the formation process in the laboratory. The liquid is Mezcal M1, from Table \ref{table:mezcales}.}
\label{fig:formation2}
\end{figure}

In this study the traditional reed was replaced by a glass pipette of  25 ml, with approximately the same exit diameter as the reed (2 mm). The sample of mezcal  (or other test liquids) of about 20 ml was placed in the pipette and was held fixed in a vertical position with a laboratory holding bracket. Below the exit of the pipette, a 10$\times$10 cm$^2$  transparent plastic container was placed, filled with the same liquid to a depth of 2 cm. The process was filmed with a high speed camera (FASTCAM-APX, Photron) using diffuse back-lighting at 1500 fps. Figure \ref{fig:formation2} shows a representative case.

For the conditions relevant to this particular setup, the jet of fluid fragments into droplets, which continuously splash against the surface of the liquid resulting in the formation of air cavities that, in turn, form bubbles that eventually rise to the surface. This process has been studied in detail, for a range of different conditions and is well understood \cite{Bin:1993,Ohl:2000,Zhu:2000}.
Figure \ref{fig:formation2} shows an image of the process of formation of pearls, considering the controlled experiment (see also Movie \#2, online submission). The impingement of the fluid stream creates a cloud of bubbles beneath the surface, which leads to the formation of surface bubbles of different sizes. However, only bubbles of a certain size (approximately of 2 mm in diameter) remain in the surface for longer times.

\subsection{Tests fluids}
The physical properties of all the test liquids are reported in Table \ref{table:mezcales}. The origin of the Mezcal samples  is also given.  According to the Mexican norm, they all come from the Oaxaca state region. Five Mezcal types were considered; only one (M1) was purposely altered to change its water or alcohol content. Also, water/ethanol mixtures were tested. It is interesting to note that Mezcal has a distinctive cultural value in rural villages in Oaxaca\cite{Gross:2014}.

The surface tension was measured with a tensiometer (DynaTester, SITA). The density was measured with a 25 ml pycnometer. The alcohol content was inferred from the density measurement, considering room temperature, $T_{room}= 23 ^o$C. The viscosity was not measured; considering the alcohol content, its value was assumed to correspond to that of a water-ethanol mixture and was obtained from tables \cite{Khattab:2012}. For a few samples, the viscosity was measured using a rheometer (TA Instruments, Discovery HR-3, parallel plates, 0.5 mm gap); a 25\% difference between measurements and tabulated values was observed. Hence, tabulated values were preferred.

\newpage
\begin{sidewaystable}
\centering
\caption{Physical properties of liquids used in the investigation.}
\label{table:mezcales}
\begin{tabular}{@{}cccccc@{}}
\hline
Fluid &  type, agave species & \begin{tabular}[c]{@{}c@{}}density\\ $\rho$, kg/m$^3$\end{tabular} & \begin{tabular}[c]{@{}c@{}}viscosity\\ $\mu$, mPa s\end{tabular} & \begin{tabular}[c]{@{}c@{}}surface tension\\ $\sigma$, mN/m\end{tabular} & \begin{tabular}[c]{@{}c@{}}ethanol content\\ \% vol.\end{tabular} \\ 
M1 ({\color{red}$\bullet$})   & Los S\'anchez, espad\'in       & 919.0                                                              & 3.07                                                             & 30.6                                                                     & 55\%                                                              \\
M2 ({\color{red}$\lozenge$})   & El Amate, espad\'in       & 914.4                                                              & 2.98                                                             & 28.7                                                                     & 48\%                                                              \\
M3a ({\color{red}$\triangledown$})  & Los peregrinos (cuerpo),  espad\'in       & 912.6                                                              & 2.66                                                             & 29.2                                                                     & 41\%                                                              \\
M3b ({\color{red}$\square$})   & Los peregrinos (punta), espad\'in       & 926.9                                                              & 3.07                                                             & 29.2                                                                     & 58\%                                                              \\
M3c ({\color{red}$\vartriangle$})  & Los peregrinos (cola), espad\'in       & 933.8                                                              & 2.34                                                             & 32.7                                                                     & 31\%                                                              \\
M4 ({\color{red}$\star$})    & Los peregrinos (MC), Madre Cuishe  & 928.3                                                              & 3.08                                                             & 29.9                                                                     & 52\%                                                              \\
M5 ({\color{cyan}$\times$})    & Los peregrinos (reposado), espad\'in  & 949.1                                                              & 2.58                                                             & 34.0                                                                     & 39\%                                                              \\
MA1 ({\color{blue}$\circ$})   & M1 + W or E, espad\'in  &  906.2 to 965.8 & 1.83 to 3.03 & 25.0 to 41.0  & 22\% to 82\%  \\
MA2  ({$\circ$})  & M1+W/E (55/45), espad\'in  & 897.4 to 901.2 & 2.82 to 2.89 & 28.4 to 29.1  & 53\% to 57\%  \\
WE  ({\color{magenta}$\square$})  & water-ethanol mixtures, - & 791.0 to 1000.0 & 1.10 to 3.01 & 23.0 to 72.0  & 0\% to 100\%  \\
\hline
\end{tabular}
\end{sidewaystable}

\subsection{Measurement of pearl lifetime}
To accurately measure the lifetime of single bubbles, a second experimental arrangement was used. A short cylindrical glass container of 1.6 cm in diameter and 1.2 cm in height was filled with the test liquid; the rim of the container was roughened to make it slightly hydrophobic. The amount of liquid was slightly larger than the volume of the container, such that a convex meniscus was formed. At the side of the container, a needle was inserted through the wall via a plastic stopper plug. Bubbles were formed by slowly pushing air through the needle with a syringe. Since the free surface was slightly convex, when a bubble reached the surface it moved to the center of the container where it could be filmed. The bubble travelled approximately 1 cm through the liquid, lasting approximately 0.03 s from its formation until it reached the surface.

Experiments were performed with needles of different gages ranging from 159 $\mu$m to 210 $\mu$m of internal diameter, to produce bubbles with slightly different diameters. The setup was illuminated from the bottom with a LED light. The bubble diameter was measured from the image obtained from the top; this measurement overestimates the bubble diameter of an equivalent spherical bubble by approximately 13\% for the highest Bond numbers\cite{Teixeira2015}. The process of rupture was measured with two synchronized high speed cameras (Phantom SpeedSense), one aligned from the top and the other from the side.  Different recording rates were considered: since the rupture time could take tens of seconds, an ordinary 30 fps was used to determine the bubble lifetime. To measure the speed of rupture of the film, the recording rate was as high as 5,000 fps. A typical high speed image sequence is shown in Fig. 1.c of the main document (see also Movie \#3, online submission).

All experiments were performed under standard laboratory conditions. The container  was thoroughly rinsed with distilled water prior to each  experiment.

\begin{figure}
\centering
\includegraphics[width=0.6\textwidth]{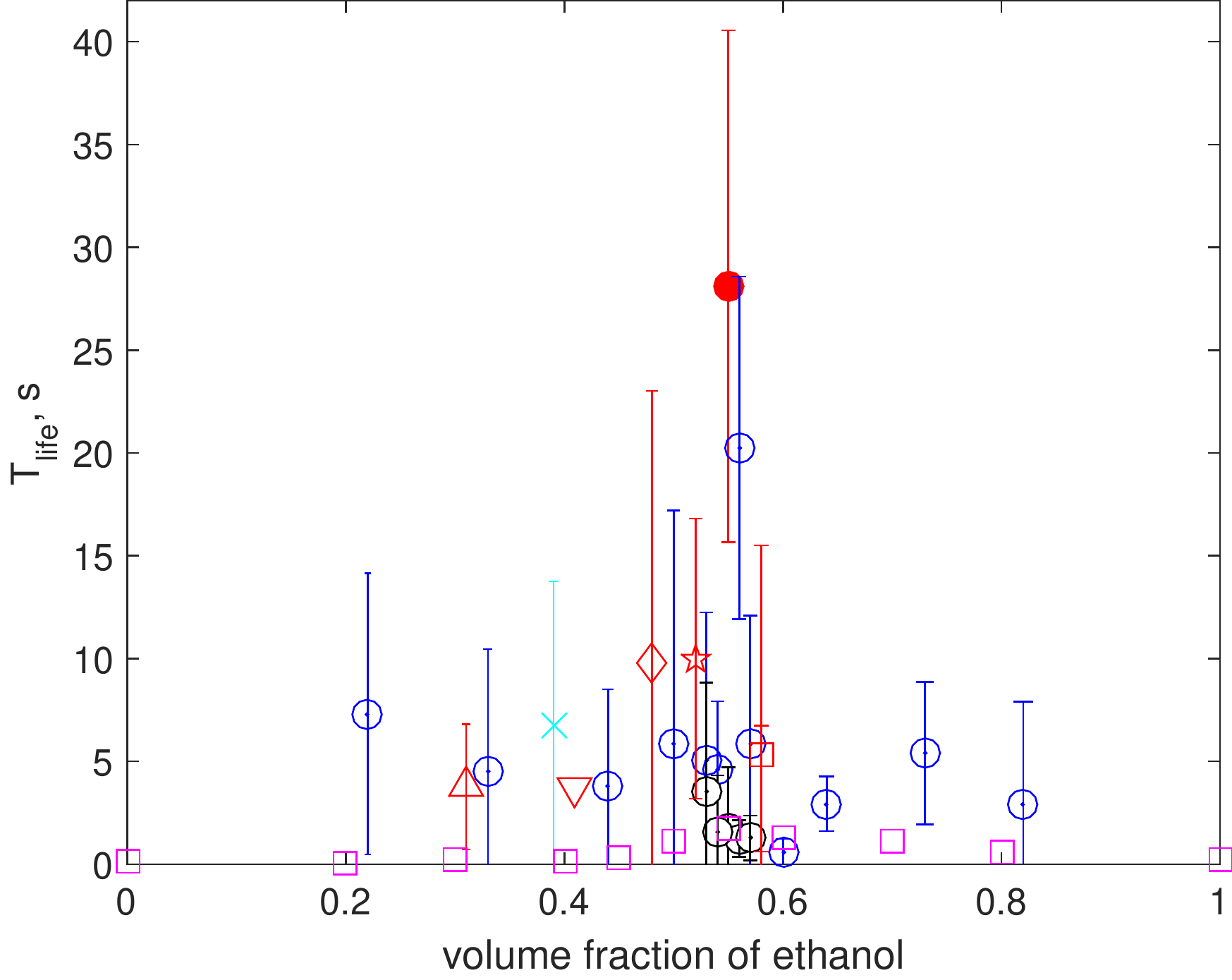}
\caption{Bubble lifetime as a function of ethanol concentration, for all test liquids. Symbols as in Table \ref{table:mezcales}. The error bars denote standard deviation of the measurements.}
\label{fig:test_all}
\end{figure}

Figure \ref{fig:test_all} shows the measured bubble lifetime, $T_{life}$, as a function of ethanol content for all the liquids used in the investigation. The symbols correspond to the nomenclature in Table \ref{table:mezcales}. One Mezcal was used to test the traditional technique (fluid M1).

\subsection{Thickness of the film during bursting}
The thickness of the film can be inferred from the speed at which the leading piercing edge moves as the bubble bursts. Considering the Taylor-Culick velocity \cite{Taylor1959,Culick1960}:
\begin{equation}\label{eqn:TCvelocity}
  V=\sqrt{\frac{2\sigma}{\rho h_{rup}}}
\end{equation}
where $\sigma$ and $\rho$ are the surface tension and liquid density, respectively. $h_{rup}$ is the thickness of the film. Experiments were conducted using fluid $M1$ and bubbles with the same diameter ($D=1.9$ mm). Five experiments were conducted, under the same nominal conditions, using a recording rate of 5000 fps. It was found that the thickness was $h_{rup}=23.5 \pm 4.8 \mu$m for the image shown Fig. (1.c) of the main document. This value is relatively large, compared to what has been measured for the case of water and seawater\cite{Lhuissier2012}. This value was used as a guideline for the numerical simulations and will be discussed in the scaling of the bubble lifetime.

\subsection{Raw data}
\begin{figure}
\centering
\subfigure[]{\includegraphics[width=0.6\textwidth]{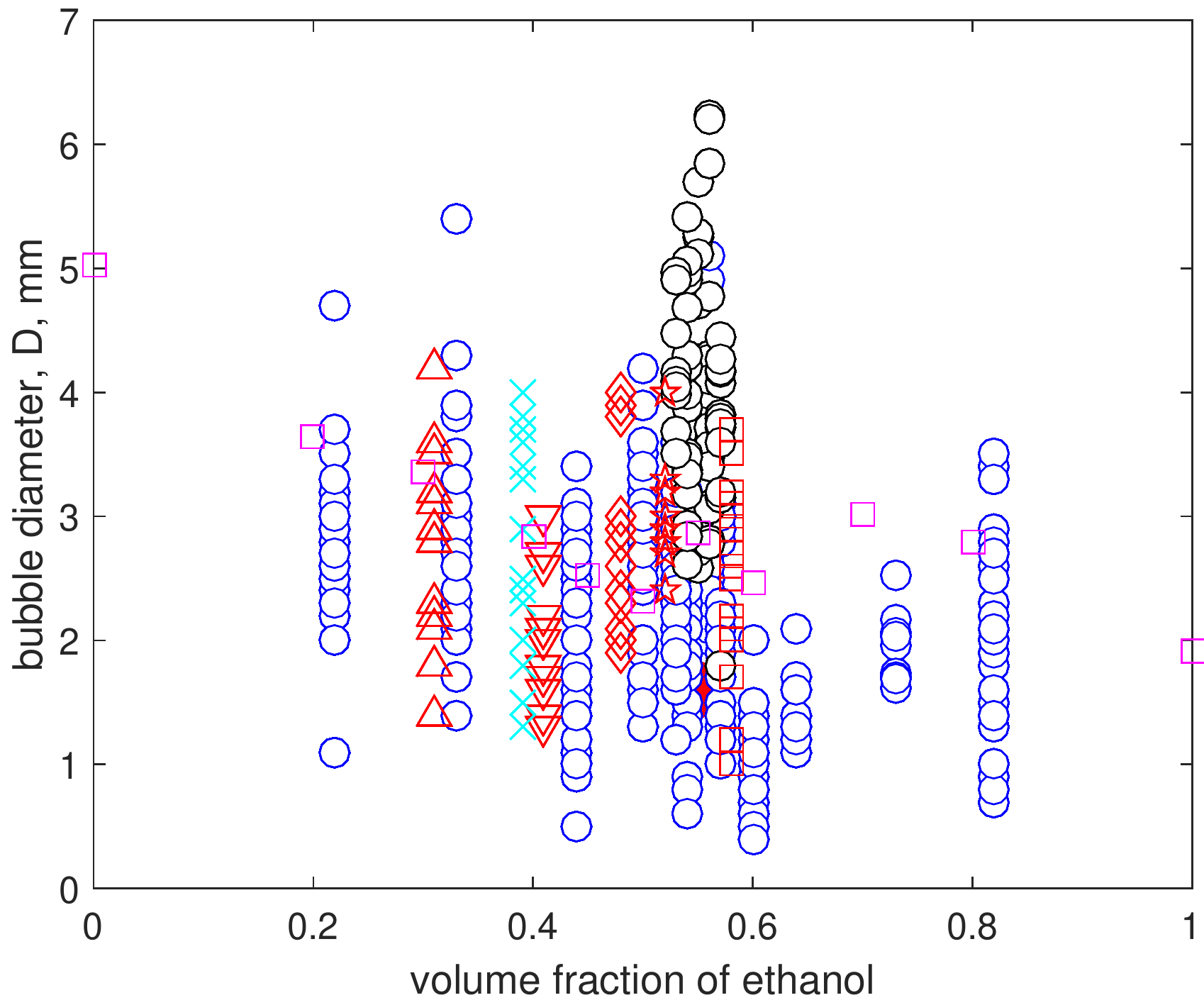}} \\
\subfigure[]{\includegraphics[width=0.6\textwidth]{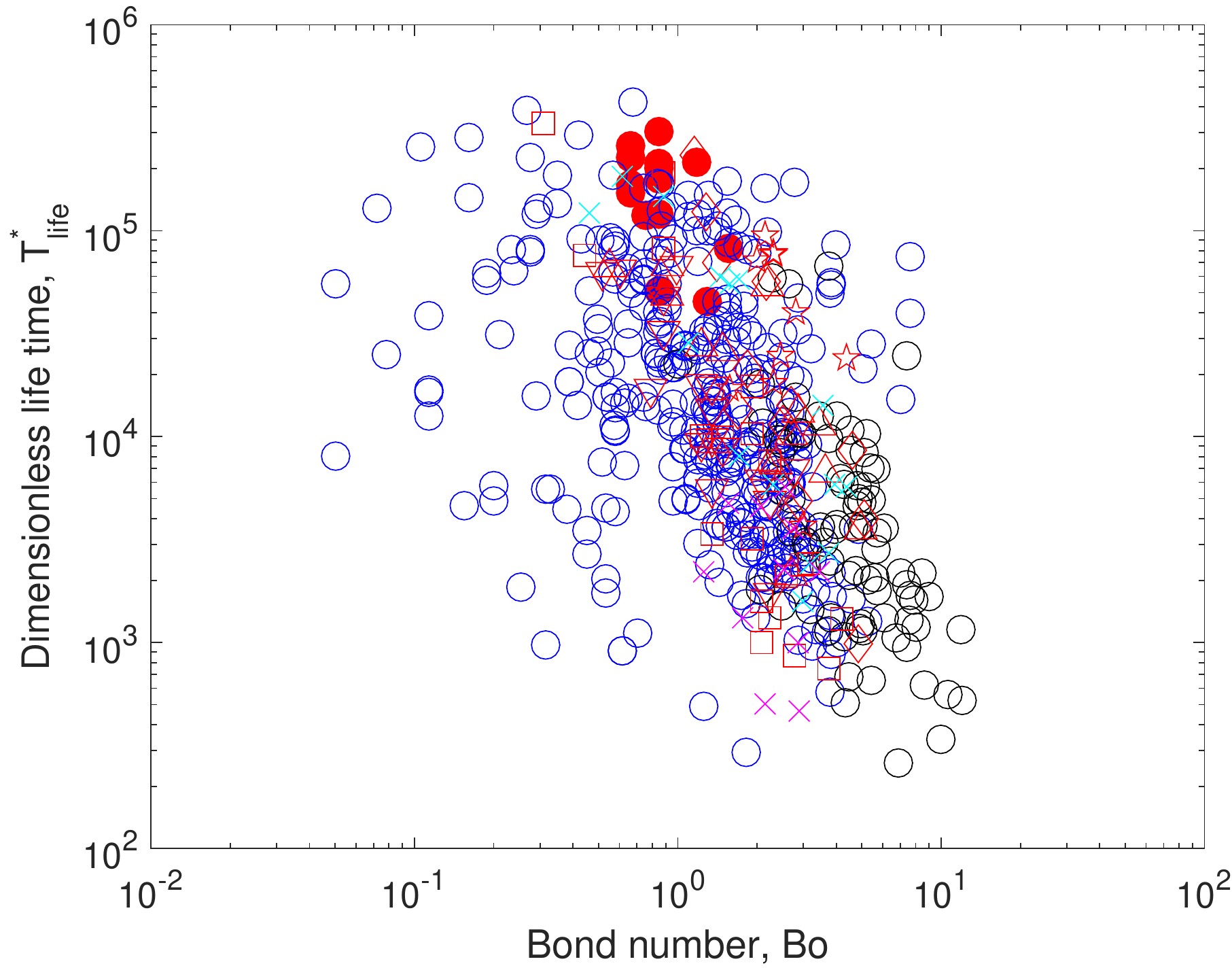}}
\caption{(a) Bubble diameter as a function of ethanol content, for all tests conducted in this investigation. (b) Dimensionless lifetime, $T^{*}_{life}$ (Eqn. \ref{eqn:dimless_time_1}) as a function of Bond number, $Bo$, for all tests conducted in this investigation. Symbols as in Table \ref{table:mezcales}. }
\label{fig:data}
\end{figure}

A minimum of 15 measurements were conducted for each of the test fluids. Figure \ref{fig:data} (a) shows the bubble diameter as a function of ethanol content for all the samples. Even for the same nominal conditions, the bubble size varied significantly. In the figures where error bars are shown, these represent the standard deviation of the measurements. In some cases, for the sake of clarity, the error bars are not shown.

Figure \ref{fig:data} (b) shows the dimensionless lifetime of bubbles as a function of Bond number (see dimensionless analysis below), for all experiments conducted in the investigation. Clearly, as in the case of the bubble diameter, the lifetime measurements show a significant scatter. We attribute this characteristic variability to the inherent nature of artisanal mezcal.

\newpage
\section{Numerical Simulations}

Direct numerical simulations were conducted by solving the Navier-Stokes equations coupled with the Level-Set method. We refer the reader to  \cite{Abadie2015,Atasi2018} for a detailed description of the method and its validation. Briefly, Navier-Stokes equations are solved for two Newtonian and incompressible fluids using the finite volume method (second order accurate in time and space). Continuity is ensured through a projection method, and the capillary contribution is considered through the classical Continuum Surface Force method. The interface position is tracked using the Level-Set method where the transport of the signed distance to the interface is controlled through the re-distancing techniques. The key point in the numerics presented here is to account for the surfactant concentration both in the liquid and on the interface as given by \cite{stone_simple_1990}:
\begin{equation}\label{Eq:bulk}
\frac{\partial C}{\partial t}+ V_c \cdot \nabla  C= \nabla \cdot \left(D_c \nabla C\right)
\end{equation}
\begin{equation}\label{Eq:surface}
\frac{\partial \Gamma}{\partial t}+ \nabla_S \cdot \left( V_S \Gamma\right)= D_s \nabla_S ^2   \Gamma+S_{ \Gamma}
\end{equation}
where $C$ is the surfactant concentration in the liquid phase, $\Gamma$ is the surfactant concentration on the gas-liquid interface, $D_c$ and $D_s$ are the diffusion coefficients of the surfactants in the liquid phase and along the interface, respectively,  $V_c$ is the velocity field of the liquid phase, $V_S$ is the projection of $V_c$ on the tangent to the interface, $\nabla_S=\left(\left(\mathbf{\bar I}-\left(n \times n \right)\right). \nabla \right)$ is the surface gradient operator and $S_{\Gamma}$ is the flux of surfactants from the liquid phase to the interface, due to the adsorption/desorption of the surfactants, i.e., $S_{\Gamma}=\left(D_c n . \nabla C \right)|_I$, where the subscript $I$ denotes the bubble-liquid interface. It is given by \cite{muradoglu_front-tracking_2008}:
\begin{equation}
S_{\Gamma}=k_a C_I \left(\Gamma_\infty- \Gamma\right)-k_d  \it\Gamma
\end{equation}
where $k_a$ and $k_d$ are adsorption and desorption kinetic constants, respectively, and $C_I$ is the surfactant concentration in the liquid in contact with the interface.

It is assumed that the surface tension depends on the surfactant concentration on the interface according to an equation of state derived from the Langmuir adsorption isotherm:
\begin{equation}\label{Eq:Lang}
\sigma=\sigma_0+R T\Gamma_\infty \ln\left(1-\frac{\Gamma}{\Gamma_{\infty}}\right)
\end{equation}
where $R$ is the ideal gas constant, $T$ is the absolute temperature, $\sigma_0$ is the surface tension of the clean interface and $\Gamma_{\infty}$ is the maximum packing concentration of surfactants on the interface.

The numerical solution of these equations is extensively described in \cite{Atasi2018}. The proper implementation of each term, in particular the surfactant transport on the gas-liquid interface and the computation of the resulting Marangoni stress,  has been verified by adapted validation cases and a free rising bubble situation was compared with results from  the literature.

\subsection{Mesh}
Simulations were performed with a non-uniform axisymmetric orthogonal mesh characterized by 400 and 200 cells in the vertical and radial direction, respectively. The mesh size was refined in the vicinity of the interface to be able to properly capture both the film drainage and its rupture. We have observed that the numerical rupture of the film occurs when the film thickness is about $h^{\rm (num)}_{rupt} \approx 5\Delta$ where $\Delta$ is the grid size. From the experiments (see Section 1.4, above), the film thickness at rupture is $h_{rupt} \approx 24 \mu$m. Due to computing limitations, the minimum grid size we could apply in the numerical simulations was $\Delta = 10 \mu$m, i.e. $h^{\rm (num)}_{rupt} \approx 50\mu$m, which should essentially explain why numerical lifetimes are shorter than experimental ones.

In the case of a film drainage between two rigid interfaces, and in the limit $Bo \ll 1$, the thinning at the apex follows $h \propto t^{-1/2}$ \cite{Scheid2012}. Consequently, a decrease by a factor two for the the critical film rupture should imply an increase by a factor four of the lifetime. Though this correction factor is certainly a good approximation and would make the numerical points overlapping the experimental points in Fig.\ref{fig:dimensionlesslife}, we did not applied this correction as the drainage dynamics with surfactants can significantly differ from the one between rigid interfaces, especially at the transition for $Bo \approx 1$ and for $Bo \ll 1$.

\subsection{Physical properties for surfactant modeling}

A certain surfactant is considered in the simulation in order to represent the effect of the interface contamination. The corresponding properties are $\Gamma_ \infty/\sigma_0 =0.1/(RT)$, where $R$ is the ideal gas constant and $T$ is the absolute temperature, $k_a = 3$ m/(mol.s), $k_d = 1$ s$^{-1}$ and $C_\infty = 25 $ mol/m$^3$, $D = 10^{-9}$ m$^2$/s and $D_s = 10^{-14}$ m$^2$/s. The fluid properties correspond to water alcohol mixtures \cite{Khattab:2012}.

\begin{figure}
\centering
\vspace{0.5cm}
\includegraphics[width=0.6\textwidth]{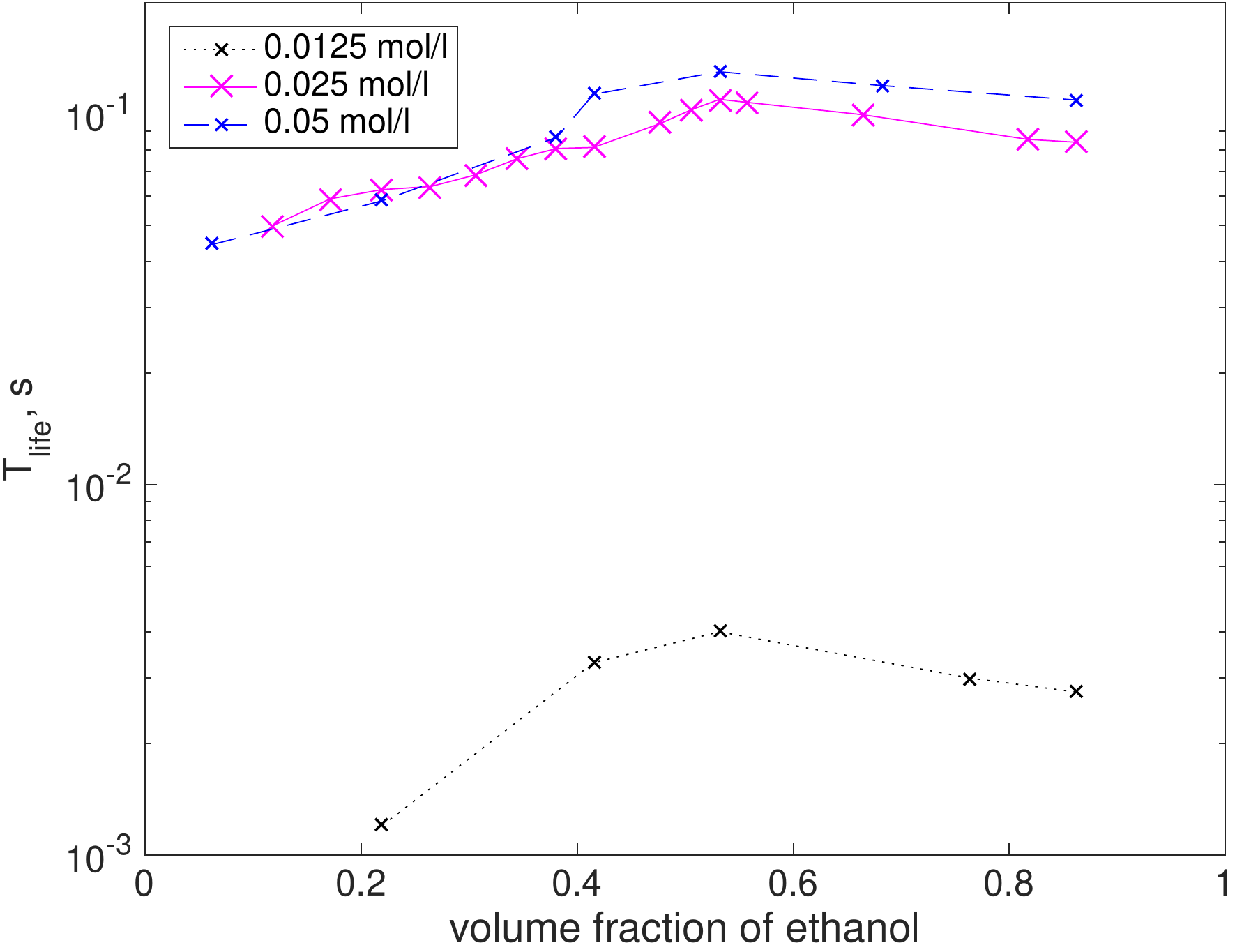}
\caption{Numerical results.  Lifetime as a function of ethanol content, for different concentrations of surfactants.}
\label{fig:numerical1}
\end{figure}

Figure \ref{fig:numerical1} shows the lifetime as a function of ethanol volume fraction, for three different amounts of surfactants. The trend is similar for all cases, but the lifetime increases with the amount of surfactants.

\subsection{Validation}
\begin{figure}
\centering
\includegraphics[width=0.95\textwidth]{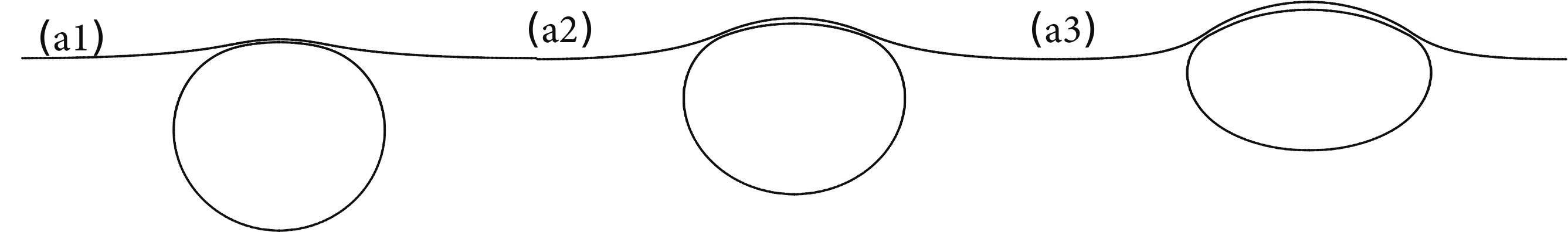}
\caption{Shape of bubbles obtained numerically for different values of $Bo$: (a1) $Bo=0.5$; (a2) $Bo=1.5$; (a3) $Bo=3.2$.}
\label{fig:numerical2}
\end{figure}

The numerical results were validated in \cite{Atasi2018}. An additional validation was conducted for this study, considering the shape of bubbles floating on the surface as the film drains. The shape that the bubble adopts while resting on the free surface changes with the value of the Bond number is shown in Fig. \ref{fig:numerical2}. The predictions are in close agreement with recent experiments\cite{Teixeira2015}. More importantly, the simulations do capture the retarded drainage due to surface tension gradients induced by concentration gradients of surfactants along the interfaces. Additionnally, the drainage dynamics associated to the deformability of the interfaces makes the film non-uniform along the bubble. In particular, a neck region occurs for sufficiently large Bond number, associated to a local thinning of the film (Fig.~\ref{fig:numerical2}a2). And the position of the neck changes position from the top of the bubble toward the liquid bath, as the Bond number increases (Fig.~\ref{fig:numerical2}a3). The appearance of the neck region has been documented for the case of films\cite{Nierstrasz1998}, but not as extensively for the case of surface bubbles\cite{Lhuissier2012}. This thinning zone leads to the rupture of the film which, as shown in Fig. (1.c), of the main document, generally does not occur at the bubble apex. The results indicate that the position where the rupture starts is a function of the Bond number.

\newpage
\section{Physical mechanisms that produce surface tension gradients}

Three physical mechanisms may induce gradients of surface tension: surfactant transport, heterogenous evaporation, and local changes in temperature. All of these mechanisms affect the time the film takes to drain and, therefore, the lifetime of the bubble. Let us first consider the effect of the concentration of alcohol. For example, the addition of 10\% of mass of ethanol to water, decreases the surface tension of the mixture more than 30\%, as seen in \cite{Vazquez1995}; these authors also showed that variations in temperature had a small effect: an increase of 30 degrees resulted in a reduction of surface tension of less than 6\%. On the other hand, Eastoe and Dalton \cite{Eastoe2000} concluded that surfactants can change the surface tension of ethanol significantly, but only when their concentration is increased orders of magnitude.

Finally, it has been shown that in ethanol/water mixtures, the alcohol evaporates at a constant rate regardless of the initial concentration \cite{Innocenzi2000}. Therefore, the evaporation rate would be uniform across the film, avoiding gradients of surface tension. However, the evaporation rate varies with the thickness of the film, as it is the case in Mezcal pearls. Such a non-uniform evaporation naturally results in the decrease of the lifetime. Hence, we do not discard the effect of evaporation, but given the long lifetimes observed experimentally, we consider this effect to be small.

In summary, we can argue that surface tension gradients result primarily from the changes in the surfactant concentration. To fully resolve this issue, the numerical scheme would have to account for both evaporation and thermal effects, but accounting here only for surfactant-induced Marangoni effect is thus qualitatively satisfactory.

\newpage

\section{Dimensional analysis}

As proposed by Barenblatt\cite{Barenblatt2003}, dimensional analysis can be used to understand a physical phenomena more deeply. There are several ways in which the relevant dimensionless numbers can be identified. We consider two methods below and then present some literature models.

\subsection{Dimensionless groups by simple inspection}
The drainage dynamics results from a balance between viscous forces and, either gravitational, or surface tension forces, or both together. The ratio that determines the relative importance between gravitational and surface tension effects is the Bond number,
\begin{equation}
    Bo = \frac{\rho g D^2}{\sigma} \label{eqn:Bo} \,,
\end{equation}
where $\rho$ is the liquid density, $g$ is the gravitational acceleration, $\sigma$ is the surface tension, and $D$ is the equivalent bubble diameter.

The viscous effects can be readily incorporated into the normalization of the lifetime, leading to:
\begin{equation}
    T_{life}^* = \frac{T_{life}\sigma}{\mu D} \label{eqn:dimless_time_1} \,,
\end{equation}
where $\mu$ is the liquid viscosity. Note that surface tension effects have been arbitrarily chosen here to scale the lifetime. An other choice built on gravitational effects would have been valid also, and as discussed below, these two time scales are identical for $Bo=1$. Either way, one can propose the following functional relationship between $T^*_{life}$ and $Bo$:
\begin{equation}\label{eqn:dimless_time_0}
 T^*_{life} = \Phi(Bo) \,.
\end{equation}

\begin{figure}
\centering
\includegraphics[width=0.6\textwidth]{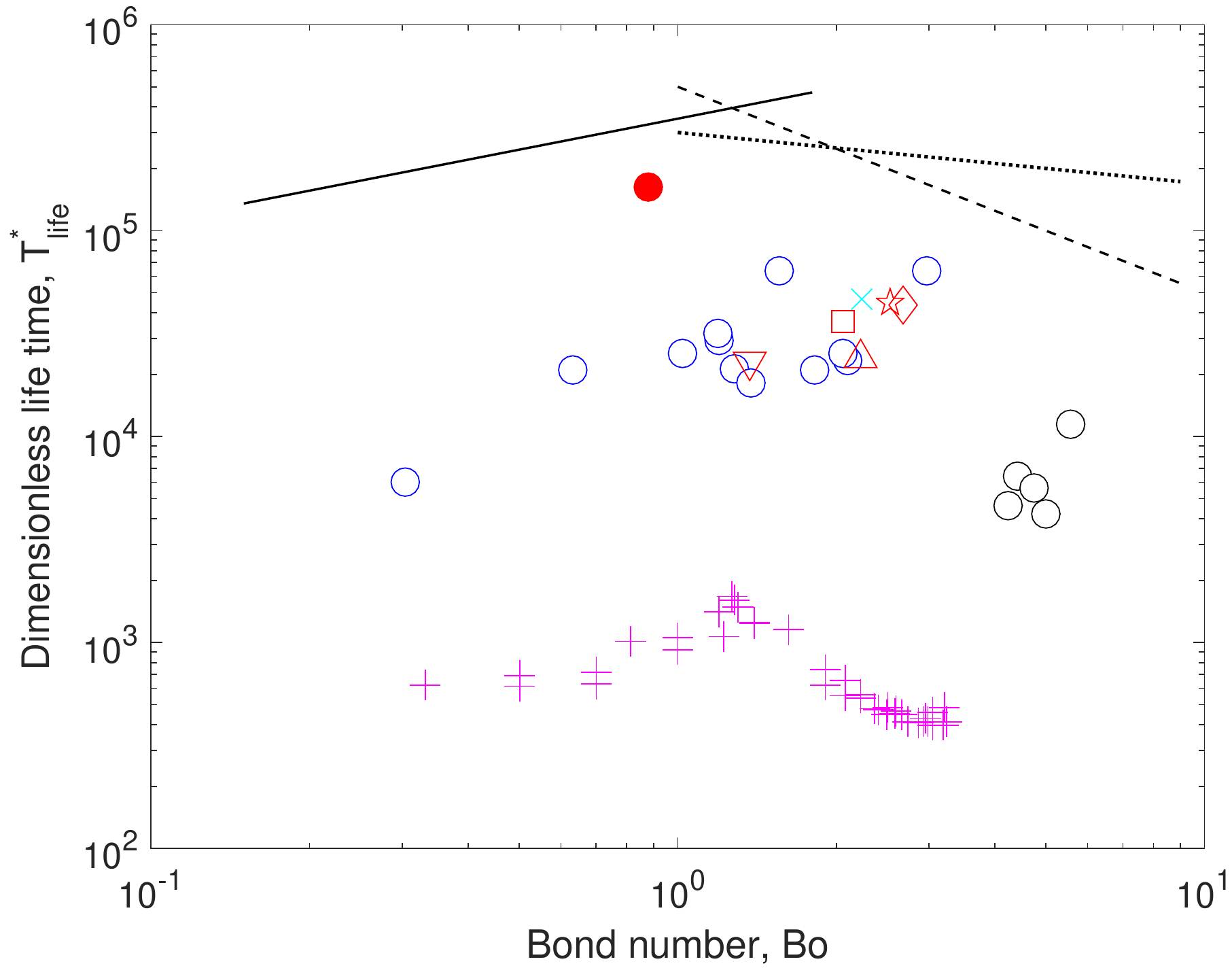}
\caption{Dimensionless lifetime as a function of Bond number. The ({\color{magenta}$+$}) symbols are the results of numerical simulations, not to scale. Symbols as in Table \ref{table:mezcales}. For clarity, error bars are omitted. The lines show trends from different predictions: solid line, from Howell \cite{Howell1999}, for small $Bo$; dashed line, Kocarkova \emph{et al.}\cite{Kocarkova2013}, for large $Bo$; and dotted line, from Lhuissier and Villermaux \cite{Lhuissier2012}, also for large $Bo$.}
\label{fig:dimensionlesslife2}
\end{figure}

Figure \ref{fig:dimensionlesslife2} shows the bubble lifetime, in terms of $T_{life}^*$, as a function of the Bond number, $Bo$, along with trends found in literature and discussed later on. A clue that the scaling by inspection has some limit is the large values of the dimensionless lifetime, namely $10^4-10^5$ time-scale units.

Now, the reason of this disagreement could be found in the presence of surfactants. If one defines $\Delta \sigma$ as the change in surface tension resulting from surfactant gradients along the interface, and the relative variation as
\begin{equation}\label{eqn:Pui}
\Pi= \frac{\Delta \sigma}{\sigma} \,,
\end{equation}
the relative variation such as one can write that
\begin{equation}\label{eqn:surfactants}
T^*_{life}= \Phi'(Bo,\Pi) \,.
\end{equation}

And as explained in the main document, we have even found that the effect can be de-correlated such that
\begin{equation}\label{eqn:surfactants2}
T^*_{life}= \Phi(Bo)\cdot\Psi(\Pi) \,.
\end{equation}

This inspection above teaches us that the surfactant-induced (Marangoni) stresses should be considered in order to determine an approrpiate time scale.

\subsection{Dimensionless groups from the analysis to the flow type}

A recent paper by Champougny {\it et al.} \cite{Champougny2016} discusses the unsolved problem of bubble drainage in presence of surfactants. They propose an ad-hoc model based on an extrapolation length $\lambda$ that allows describing the transition from stress-free interfaces ($\lambda=\infty$) to no-slip interfaces ($\lambda=0$). The flow between stress-free interfaces is a plug flow at leading-order, dominated by extensional viscous stresses. Debregeas {\it et al.} \cite{Debregeas1998} have shown that the film thickness of such a ``bare" bubble decays exponentially in time and that the film always ruptures at the apex. Contrarily, the flow between no-slip (rigid) interfaces is a Poiseuille-like flow dominated by shear viscous stresses (as described fo instance in \cite{Bhamla2014}). Now, surfactant-induced Marangoni stresses have been shown to rigidifies, at least partially, the interfaces, such as the flow is essentially a shear flow (see for instance Champougny {\it et al.} \cite{Champougny2015} for film formation or Atasi {\it et al.} \cite{Atasi2018} for confined bubbles in microchannels, both in presence of surfactants). Even traces of surfactants, such as impurities, have been found to induce dominant Marangoni stresses \cite{Ratulowski89}. Additionally, local thinning mechanisms of the film have been observed in soap bubbles \cite{Lhuissier2012}, reminiscent to the marginal regeneration \cite{Mysels1959}, and eventually leading to the rupture of the film away from the apex.

As already mentioned, enough evidences in the present study plead for a flow dominated by viscous shear stresses, such as the long lifetime as compared to the extensional timescale (see Fig.~\ref{fig:dimensionlesslife2}) or the film rupture observed at other locations than the apex. The hypothesis of a dominant shear flow has also been corroborated by the simulations presented in Fig.~4 of the main paper, where the surface velocity is shown to be drastically reduced in presence of surfactants, as if the interfaces were partially rigidified.

In absence of a simple model available that accounts for the presence of surfactants and with the aim to find an appropriate time scale, we will consider in the following a drainage dominated by a shear viscous flow and balanced either by gravity or capillarity. Starting with a gravity-driven drainage, the momentum balance at leading order in the lubrication approximation writes
\begin{equation}
\mu \partial_{yy}u = \rho g\,,
\end{equation}
where the cross-stream coordinate $y$ in the film can be scaled with the critical thickness for rupture $h_{rup}$.  The choice of this length scale is justified as the drainage dynamics usually `forgets' about the initial condition since the main contribution to the lifetime occurs at the later stage, i.e. when the film is the thinnest, henceforth the viscous dissipation the highest. However, as indicated above, $h_{rup}$ was only measured for the case of pure mezcal (M1 fluid, from Table 1), and not for all experiments conducted here. In Fig. 3, we have thus taken the value of $h_{rup}$ constant and equal to the mean measured value of 24 $\mu$m as explained in section 1.4. For consistency, the same value has also been taken in the simulations (see Section 2.1).

By estimating that $u\sim D/t_g$, one can obtain the timescale for drainage driven by gravity, i.e.,
\begin{equation}
t_g = \frac{\mu D}{\rho g h_{rup}^2} \,.  \label{eqn:tg}
\end{equation}
Similarly, the capillary-driven drainage results from the balance
\begin{equation}
\mu \partial_{yy}u = \partial_x P \,,
\end{equation}
where the streamwise coordinate $x$ can be scaled by $D$ and where the pressure $P$ scales with the Laplace pressure of the order of $\sigma/D$. By estimating that $u\sim D/t_c$, the timescale for drainage driven by capillarity is
\begin{equation}
t_c = \frac{\mu D^3}{\sigma h_{rup}^2} \,.   \label{eqn:tc}
\end{equation}
Comparing these two timescales leads to the Bond number, namely $Bo = t_c/t_g$, but it does not have the same meaning, dynamically speaking.

Indeed, it is interesting to note that the Bond number plays two roles in this problem, a static and a dynamic one. On the one hand, it allows to evaluate the static shape of the interfaces, namely a spherical bubble underneath an almost undeformed liquid surface for $Bo \ll 1$, and a deformed bubble under a deformed liquid surface for $Bo \gg 1$. On the other hand, as explained above, it allows to evaluate the dominant driving force for drainage, namely the capillary force for $Bo \ll 1$ and the gravity force for $Bo \gg 1$. The striking feature of the present problem is that $Bo\sim 1$ indicates a transition, which coincides with the maximum bubble lifetime.

The dimensionless lifetime can now be defined in terms of the capillary time scale, $t_c$ for instance:
\begin{equation}
  {T^{**}_{life}} = \frac{h_{rup}^2}{D^2}  \frac{T_{life}\sigma}{\mu D} =  \varepsilon^2 {T^{*}_{life}}  \label{eqn:dimless_time_3}\\
\end{equation}
where Eqn. (\ref{eqn:dimless_time_1}) has been used, and where the aspect ratio $\varepsilon=h_{rup}/D$ has been introduced. This aspect ratio expresses that the viscous shear forces act perpendicularly to the capillary/gravity forces. Not that extensional forces in the case of an extensional flow acts along the same direction that for the forces responsible for drainage, such as $\varepsilon$ can be set to unity. Consequently, $T^*_{life}$ is the appropriate time scale for an extensional drainage. In the present problem, $\varepsilon^2 \sim 10^{-4}$, explaining why the dimensionless lifetime $T^*_{life}$ is so large in Fig.~\ref{fig:dimensionlesslife2} and why $T^{**}_{life}$ as been used instead in Fig.\ref{fig:dimensionlesslife}.

As expected the data shows a clear transition in trend at around $Bo\approx1$. One can now use the timescale for $T^{**}_{life}$ to derive the trends in both limits of large and small Bond numbers. In the limit of large Bond number, the drainage is governed by gravity and  using Eqn.~(\ref{eqn:tg}), one can write
\begin{equation} \label{eqn:Bogg1}
T^{**}_{life} \propto Bo^{-1} \qquad (Bo \gg 1)  \,,
\end{equation}
while in the limit of small Bond number, the drainage is dominated by capillarity and using Eqn.~(\ref{eqn:tc}), one can write
\begin{equation} \label{eqn:Boll1}
T^{**}_{life} \propto 1 \qquad\qquad\!\! (Bo \ll 1) \,.
\end{equation}
The trends above are also plotted in Fig. \ref{fig:dimensionlesslife}. They are only there to qualitatively guide the eye  as the experimental and numerical points do not cover more than one decade and lie precisely on the transition at around $Bo=1$, rendering any quantitative comparison meaningless.

\subsection{Literature models}

\subsubsection{$Bo \gg 1$}

In the limit of Large Bond numbers, the bubbles are large and the film near the apex is essentially uniform with a drainage driven by gravity.

In the case of rigid interfaces, an analytical solution exists for the dimensionless lifetime as follows \cite{Scheid2012}
\begin{equation}\label{eqn:scheid}
  T^{**}_{life} = \frac{3}{2} Bo^{-1}.
\end{equation}
Notably, this prediction underestimates most of the experimental points in Fig. \ref{fig:formation}, which indicates that the Marangoni stress in the experiments is large enough to retard the drainage longer than what rigid interfaces would do. This is only possible if the mean flow is negative, i.e. towards the apex, entrained by Marangoni stresses in the same direction.

In the case of stress-free interfaces, Kocarkova \emph{et al.}\cite{Kocarkova2013} calculated the evolution of the film thickness considering an extensional flow in the film, assumed to be uniform and axisymmetric. They observed an exponential thinning, from which the dimensionless lifetime can be shown to be:
\begin{equation} \label{eqn:t_b_Kocarkova}
  T^*_{life} \sim Bo^{-1} \,,
\end{equation}
using the extensional timescale.
The trend given by (\ref{eqn:t_b_Kocarkova}) is shown by the dashed line in Fig.~\ref{fig:dimensionlesslife2}, even though it predicts much shorter lifetime than those observed experimentally.

Champougny \emph{et al.}\cite{Champougny2016} extended the analysis of \cite{Kocarkova2013} considering certain levels of surface rigidification of the interfaces to account for the presence of surfactants. They also found an exponentially thinning of the film with time, which leads to the same functional relation of Eqn. (\ref{eqn:t_b_Kocarkova}) but the proportionality constant was lager in the case of surfactants. Interestingly, they  also found that the puncture of the bubble film changed from the apex to the foot as the amount of surfactants increased.

Finally, Lhuissier and Villermaux \cite{Lhuissier2012} considered the case of bubbles in water. They argued that the viscous draining scaling was not appropriate for water due to the phenomenon known as marginal regeneration already mentioned in the previous section\cite{Nierstrasz1998}. In short, the balance of capillary pressure from the film curvature, the meniscus at the foot of the bubble and surface tension gradients lead to a non uniform film thickness which causes the appearance of a localized pinching. Moreover, Lhuissier and Villermaux \cite{Lhuissier2012} recognized that the film breakup could also be influenced by the B\'enard-Marangoni convection flows within the film. Considering the fluctuations from the marginal regeneration convection cells and  probabilistic arguments of the puncture breakup mechanism, they found
\begin{equation}\label{eqn:t_b_Lhu1}
  T^*_{life} \sim  Bo^{-1/4}\,,
\end{equation}
which also indicates a decreasing time with an increase of $Bo$. This trend is shown by the dotted line in Fig.~\ref{fig:dimensionlesslife2}. Given the experimental uncertainty, it is hard to discern to which trend for large $Bo$ the data is closest to, but in either case it is clear that the lifetime is inversly proportional to $Bo$ when $Bo$ is larger than unity, in agreement with the scaling result (Eqn. \ref{eqn:Bogg1}).

\subsubsection{$Bo \ll 1$}

In the limit of small Bond numbers, no simple model exists for the bubble lifetime as for $Bo \ll 1$ and in the limit of rigid interfaces. This is because for small Bond numbers, the drainage is driven by capillary forces that induce surface deformations. One can only note the model of Howell \cite{Howell1999}, derived in the case of an extensional flow, and expressed as
\begin{equation}\label{eqn:t_b_Howell}
  T^*_{life} = \kappa_1 Bo^{1/2} \qquad (Bo \ll 1)\,,
\end{equation}
where $\kappa_1$ is a constant that depends on the initial and rupture thicknesses. This expression indicates that the rupture time increases with Bo: the larger the bubble, the longer it will take to burst. This calculation assumes that the film drains uniformly and axisymmetrically without surfactants. The small $Bo$ limit indicates that the bubble is nearly spherical and is mostly immersed below the liquid free surface. The trend given by (\ref{eqn:t_b_Howell}) is shown by the solid line in Fig.~\ref{fig:dimensionlesslife2} and differs from the scaling result (Eqn. \ref{eqn:Boll1}) obtained for a shear flow.

\subsubsection{$Bo \sim 1$}

Clearly, the trends are opposite for small and large Bond numbers indicating a transition at a critical value of $Bo$ around unity, where capillary and gravity forces are of comparable magnitude to drive the drainage. And this transition should correspond to a maximum lifetime, as suggested by the cross-over between the models given above. The experimental data also seem to capture the transition, despite their wide dispersion. Nevertheless, the numerical results clearly shows this maximum of the dimensionless lifetime at  $Bo\approx1$ as it was also observed for Mezcal bubbles.

\newpage


\end{document}